\documentclass[10pt,conference,a4paper]{elektron}

\usepackage[latin1]{inputenc}
\usepackage[compress]{cite}
\usepackage{graphicx}
\usepackage{subcaption}
\usepackage{float}
\usepackage{algpseudocode}
\usepackage{algorithm}
\usepackage{hyperref}
\usepackage{amssymb}
\usepackage{amsmath}
\usepackage{cleveref}
\usepackage{booktabs}
\usepackage{stfloats}
\usepackage{makecell}
\usepackage{amsthm}

\theoremstyle{definition}
\newtheorem{exmp}{Example}

\begin{document}

\title{
  Deep Recurrent Learning for Heart Sounds Segmentation based on Instantaneous Frequency Features \\
  \vspace{0.5cm}
  \large Aprendizaje profundo y recurrente para la segmentaci\'{o}n de sonidos card\'{i}acos basado en caracter\'{i}sticas de frecuencia instant\'{a}nea
}

\author{
  \IEEEauthorblockN{
    \'{A}lvaro Joaqu\'{i}n Gaona\IEEEauthorrefmark{1}$^1$,
    Pedro David Arini\IEEEauthorrefmark{1}\IEEEauthorrefmark{2}$^2$
    \vspace{0.2cm}
  }
  \IEEEauthorblockA{
    \IEEEauthorrefmark{1}\emph{Facultad de Ingenier\'{i}a, Universidad de Buenos Aires,}\\
    \emph{Instituto de Ingenier\'{i}a Biom\'{e}dica, (IIBM)}\\
    \emph{Avenida Paseo Col\'{o}n 850, C1063ACV, Buenos Aires, Argentina}
  }
  \IEEEauthorblockA{$^1$\texttt{\small{agaona@fi.uba.ar}}}
  \IEEEauthorblockA{
    \IEEEauthorrefmark{2}
    \emph{Instituto Argentino de Matem\'{a}tica "Alberto P. Calder\'{o}n", CONICET}\\
    \emph{Saavedra 15, C1083ACA, Buenos Aires, Argentina}
  }
  \IEEEauthorblockA{$^2$\texttt{\small{pedro.arini@conicet.gov.ar}}}
}

\maketitle
\thispagestyle{empty}
\pagestyle{empty}

\begin{abstract}
  In this work, a novel stack of well-known technologies is presented to determine an automatic method to segment the heart sounds in a phonocardiogram (PCG). 
  We will show a deep recurrent neural network (DRNN) capable of segmenting a PCG into its main components and a very specific way of extracting instantaneous frequency that will play an important role in the training and testing of the proposed model.
  More specifically, it involves a Long Short-Term Memory (LSTM) neural network accompanied by the Fourier Synchrosqueezed Transform (FSST) used to extract instantaneous time-frequency features from a PCG.
  The present approach was tested on heart sound signals longer than 5 seconds and shorter than 35 seconds from freely-available databases.
  This approach proved that, with a relatively small architecture, a small set of data, and the right features, this method achieved an almost state-of-the-art performance, showing an average sensitivity of 89.5\%,
  an average positive predictive value of 89.3\% and an average accuracy of 91.3\%.
  \\\\
  $~~$\emph{Resumen---}En este trabajo se presenta un conjunto de t\'{e}cnicas bien conocidas definiendo un m\'{e}todo autom\'{a}tico para determinar los sonidos fundamentales en un
  fonocardiograma (PCG).
  Mostraremos una red neuronal recurrente capaz de segmentar segmentar un fonocardiograma en sus principales componentes, y una forma muy espec\'{i}fica de extraer frecuencias instant\'{a}neas que jugar\'{a}n un importante rol en el entrenamiento y validaci\'{o}n del modelo propuesto.
  M\'{a}s espec\'{i}ficamente, el m\'{e}todo propuesto involucra una red neuronal \textit{Long Short-Term Memory} (LSTM) acompa\~{n}ada de la Transformada Sincronizada de Fourier (FSST) usada para extraer atributos en tiempo-frecuencia en un PCG.
  El presente enfoque fue evaluado con se\~{n}ales de fonocardiogramas mayores a 5 segundos y menores a 35 segundos de duraci\'{o}n extra\'{i}dos de bases de datos p\'{u}blicas.
  Se demostr\'{o}, que con una arquitectura relativamente peque\~{n}a, un conjunto de datos acotado y una buena elecci\'{o}n de las caracter\'{i}sticas, este m\'{e}todo alcanza una eficacia cercana a la del estado del arte, con una sensitividad promedio de 89.5\%,  una precisi\'{o}n promedio de 89.3\% y una exactitud promedio de 91.3\%.
  \\\\
  Palabras clave: fonocardiograma;
  transformada sincronizada de fourier;
  \textit{long short-term memory}.
\end{abstract}

\IEEEpeerreviewmaketitle

\section{Introduction}\label{sec:introduction}

Phonocardiography is a method to record the acoustic phenomena of the heart graphically.
It is used to provide information about the cardiac cycle by plotting sounds and murmurs of the heart.
The sounds result from the closure of the heart valves, and it is possible to identify at least two sounds.
The first one, $S_1$, corresponds to the closure of the atrioventricular valves (mitral and tricuspid valve) at the
beginning of the systole.
At this point, the ventricles filled with blood from the atriums and muscle contractions begin to eject the oxygenated and deoxygenated blood to the pulmonary and systemic circuits respectively.
After most of the blood has been ejected from the ventricles, the aortic and pulmonary valves close producing the
second sound, $S_2$.
Additionally, two other segments of the phonocardiogram (PCG) can be identified.
The first one is the segment $S_1$-$S_2$ called isovolumetric contraction and the second one is the segment.
$S_2$-$S_1$ called isovolumetric relaxation, which usually is shorter than the first segment.
Heart sounds segmentation dates back to 1997 where H. Liang \textit{et al.} used a deterministic algorithm based
on the normalized average Shannon energy of a PCG signal achieving a 93\% correct ratio.
This approach has, however, some drawbacks such as corrupting noise.
In the same year, H. Liang \textit{et al.} \cite{757028} proposed an algorithm based on wavelet decomposition and
reconstruction performing correctly in over 93\% of cases.
Heart sounds segmentation boomed in 2010 when Schmidt \textit{et al.} \cite{004} proposed a Hidden Markov Model (HMM)
based on time-duration called Dependent-duration Hidden Markov Model (DHMM).
Additionally, it introduced the use of annotations derived from the EKG to label training sets to train the proposed model, later used by Springer \textit{et al.} \cite{7234876} to go even further and outperform the
previous work by adding logistic regression and modifying the implementation of the Viterbi algorithm.
In 2018, Renna \textit{et al.} in \cite{8620278} have used Deep learning techniques to segment the PCG.
Their approach, motivated by a novel convolutional neural network called
U-net~\cite{DBLP:journals/corr/RonnebergerFB15} for neuronal structures segmentation in electron microscopic stacks,
used Schmidt and Springer techniques such as labelling and feature extraction (Homomorphic envelogram, Hilbert envelope,
Wavelet envelope and Power Spectral Density envelope) to outperform what was at that time known as a state-of-the-art
technique.

In this work, we propose an implementation based on a DRNN to segment PCG signals into their main components. The proposed method involves a Long Short-Term Memory (LSTM) neural network accompanied by the Fourier Synchrosqueezed Transform (FSST) used to extract instantaneous time-frequency features in
a PCG.

\section{Material and methods}\label{sec:material-and-methods}

Most of the work begins at deciding which approach and what data is going to be used to achieve the desired goal.

Deep Learning has proved to the world that it is a compelling strategy to address various kind of problems and signal segmentation is no exception.
Moreover, deep learning techniques have not been used in PCG as it has been used in electrocardiogram (EKG) signals lately.
Long Short Term Memory neurons are pretty powerful neural network units that can achieve great accuracy when trained on
time-series data. 
This data must be well acquired and processed before feeding it into the neural network.
Both stages, acquisition and processing, require a right amount of effort to execute and, most of the times, much more than training the network.

Lastly, the obtained model and results must be interpreted, verified and validated by following different techniques such as Cross-Validation (CV),
Receiver Operation Characteristic (ROC) curve and other performance metrics.
A decent examination of the recently referenced techniques will allow selecting a model that is well fitted to segment the PCG effectively.

\subsection{Database}\label{subsec:database}

The PCG recordings, extracted from the so called PhysioNet Database \cite{PhysioNet}, was utilized.

The Challenge 2016 carried out by the Computing of Cardiology (CinC) 2016 provided participants with a reasonably big dataset comprised of PCG and EKG recordings along with various annotations such as a patient identifier, abnormality of the signal and so on. Nevertheless, this challenge asked participants to classify PCG according to the pathology presented in them. Additionally, the trial provided a link to Springer's implementation of
\textit{Logistic Regression Hidden Semi-Markov Model} (LR-HSMM) for heart sounds segmentation and with it, 792 PCG recordings from 37 patients with R wave and end of T wave annotations. These wave annotations have correspondences in time with S1 and S2 in the PCG respectively, according to \cite{004}, and Springer \textit{et al.} implemented a labelling algorithm in \cite{7234876} to automatically generate labels used for training models. Furthermore, experts identified these recordings as healthy and unhealthy. The dataset maintains an equilibrium between normal and abnormal signals which is essential for training models for them to learn different features or patterns off of the data.

\subsection{Annotations and labeling}\label{subsec:annotations}

In \textit{Supervised Learning} extracting labels is a crucial step to go through. 
It can be performed manually or automatically.
The former is commonly done by specialists in the field, and the later is fundamentally an algorithm responsible for computing them.

\begin{figure}[H]
    \centering
    \includegraphics[scale=0.185]{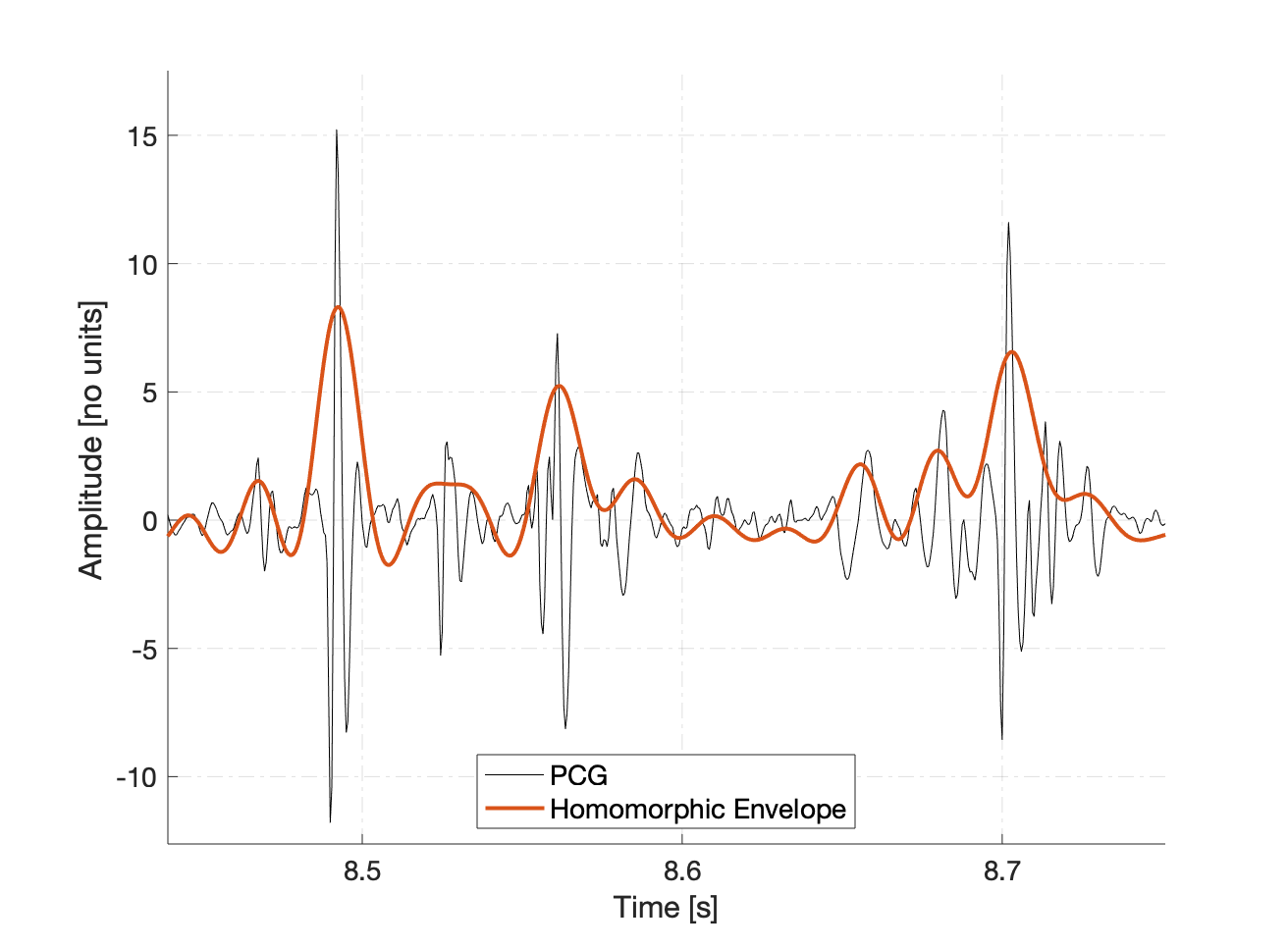}
    \caption{Homomorphic envelope (normalized) used for PCG labeling. In red, a smoother signal, corresponding the envelope, and in black, the PCG. }
    \label{fig:henv}
\end{figure}

In this work, labels were extracted using a labelling algorithm provided by Springer \cite{7234876}.
This algorithm leverages annotations provided in the dataset corresponding to the EKG waves (R-wave and end of T-wave) and the homomorphic envelope \cite{004} which is depicted in Figure \ref{fig:henv}.
Based on each annotation, a lower and upper bound is defined to label S1 and S2.
Between S2 and the S1 from the following cardiac cycle lies the diastolic interval and the is assumed to be the systolic interval.

\begin{figure}[H]
    \centering
    \includegraphics[scale=0.21]{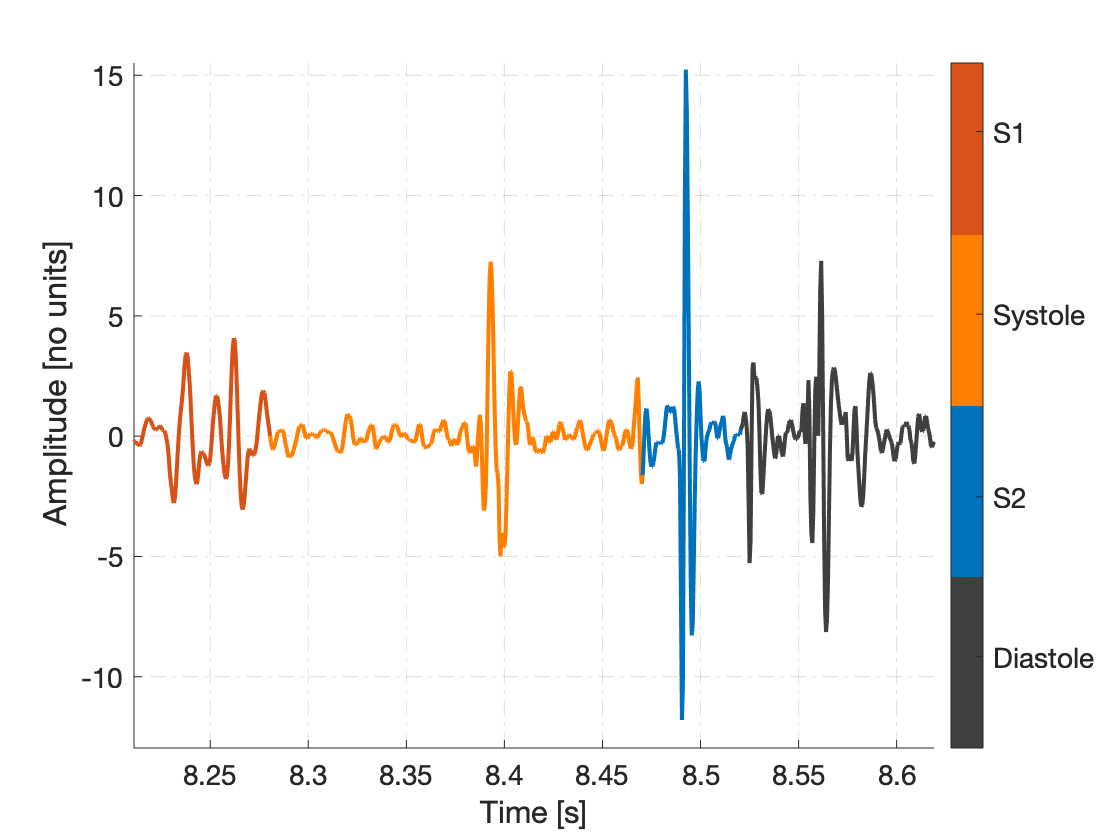}
    \caption{Automatically labeled PCG (normalized). Four states are identified in four different colors, showing the beginning and ending of each one.}
    \label{fig:labeled-pcg}
\end{figure}

It is worth mentioning that the previously mentioned algorithm should only be used offline. The training labels this algorithm yields, should only be used to train the neural network. It is not advisable to use this algorithm in real-time applications due to the dependency of the EKG signal. If so, performing PCG segmentation online could be quite troublesome. Moreover, the algorithm has to be tuned manually to retrieve reliable labels. Thus, the need to develop a method independent of EKG signals, and a trained deep neural network is a good way of solving this problem. Additionally, it can be implemented in real-time embedded systems with special care to perform an online segmentation, if that would be goal. An example of a automatically labelled PCG signal is illustrated in Figure \ref{fig:labeled-pcg}

\section{Approach and DRNN models}

\begin{figure}[H]
    \centering
    \includegraphics[scale=0.45]{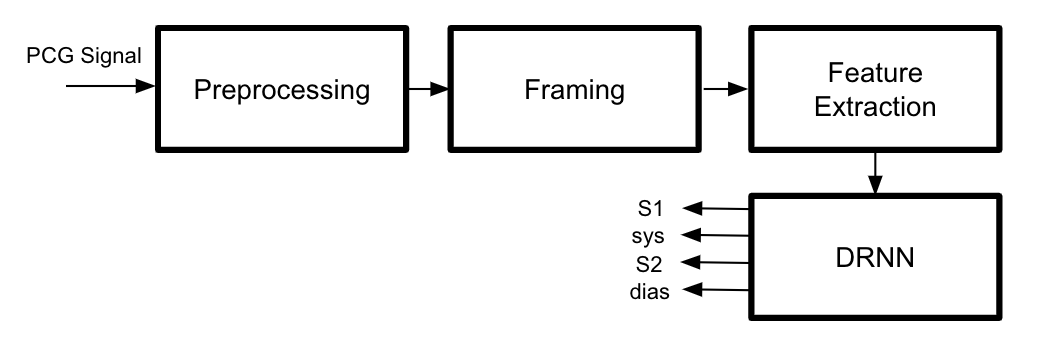}
    \caption{Heart sounds segmentation approach.}
    \label{fig:workflow}
\end{figure}

\subsection{Preprocessing}\label{subsec:signal-normalization}

Neural networks cannot ingest data without being adequately transformed. Otherwise, these would not perform at their best in both training and testing stages.
Classification techniques expect classifying observations into the right class by observing explanatory variables or features.
Most of the times, these features are not on the same scale. Thus, \textit{standarization} is an excellent technique to perform on the data before feeding it into the network.

\begin{figure}[H]
    \centering
    \includegraphics[scale=0.215]{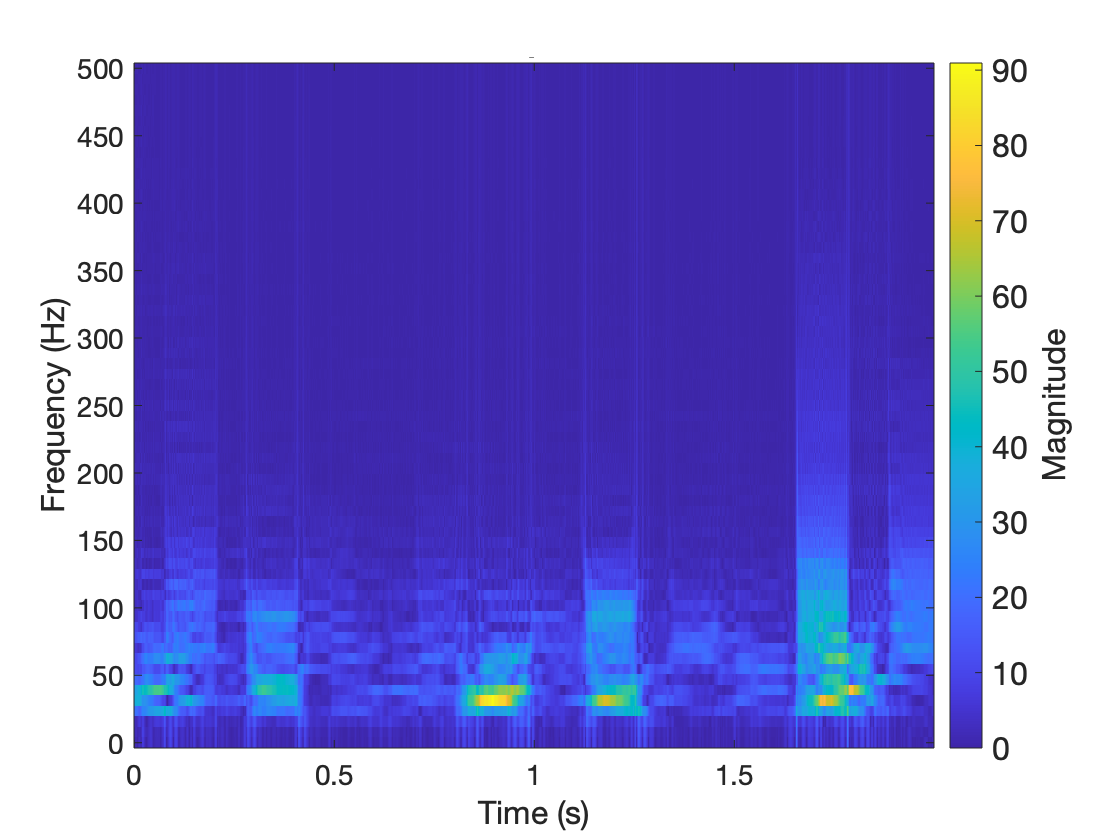}
    \caption{Fourier Synchrosqueezed Transform across a portion of a PCG. It is possible to see the two main sounds (depicted in green), and both systole and diastole intervals, associated with higher and smaller energy levels, respectively.}
    \label{fig:spectrum}
\end{figure}

\begin{equation}
    Z_i = \frac{X_i - \mu_i}{\sigma_i}, \hspace{2mm} i = 1, \dots, p
\end{equation}

Where $p$ is the number of features. Secondly, the frequency content in a PCG signal has been determined to be between 25 - 400 Hz in Schmidt work. However, we have identified that most of the frequency energy needed is contained in the 20 - 200 Hz range, as it is show in Figure \ref{fig:spectrum}.





\subsection{Framing}\label{subsec:framing}

Due to Deep Neural Networks (DNN) having a fixed input length and the signals not having the same duration, it is essential to perform a framing process on them.

N-dimensional patches were extracted from an N-dimensional signal $\boldsymbol{x}_k \in \mathbb{R}^{T}$ with a given stride $\tau$ for $k = 0, 1, \ldots, T - 1$.
These patches $\boldsymbol{z}_i$ are the inputs which the network is fed with.
Thus, $\boldsymbol{z}_i \in \mathbb{R}^{L}$ is constructed by computing \Cref{eq:patches}.

\begin{equation} \label{eq:patches}
    \boldsymbol{z}_i = \begin{bmatrix} \boldsymbol{x}_{i \cdot \tau} \\ \vdots \\ \boldsymbol{x}_{i \cdot \tau + L - 1} \end{bmatrix}
\end{equation}

Where $N = \lfloor \frac{T - 1 - L}{\tau} \rfloor$ and $L$ is the fixed length of the patches and $i = 0, 1 \ldots, N$ and $\lfloor x \rfloor$ denotes the greatest integer lower or equal than $x$.
It is worth considering that after you select $L$ you cannot frame signals whose length is lower than the selected length, therefore those signals will not be taken into account. Therefore, a good choice of $L$ is equal to the lowest signal length.

\begin{exmp}
    \textit{We want to create patches of length $L = 2000$ from a signal whose length is $T = 35000$. After these two parameters are set up, we just need to define how much overlapping we want between patches, defined by parameter $\tau$. This means if we set $\tau = 1000$, 50\% of the samples in each patch will be repeated in the adjacent ones and if it is set to $\tau = 2000$, patches will not intersect. Suppose we choose $\tau = 1000$, then $N$ is equal to 32, meaning that the signal will be framed into 32 patches. It is important to note that if $T - 1 - L$ is not multiple of $\tau$ some samples will be discarded due to the floor function $\lfloor \cdot \rfloor$}.
\end{exmp}

\subsection{Feature extraction}\label{subsec:feature-extraction}

Many features have been used to perform PCG segmentation.
Mostly envelopes were used by Schmidt \cite{004} and Springer \cite{7234876}.
The former used the so-called Homomorphic Envelope, and the latter added three more envelopes, such as the Discrete Wavelet (DWT) Envelope, the Power Spectral Density (PSD) Envelope, and the Hilbert Envelope.

Nevertheless, in this work, we proposed a different approach to accomplish PCG segmentation.
Fourier Synchrosqueezed Transform (FSST) \cite{fsst} is a technique based on Short-time Fourier Transform (STFT) that maps STFT frequencies into instantaneous frequencies of the signal at a given time $t$.
After computing the FSST on a PCG signal, just a frequency range is extracted from it. 
Frequencies in the range of 20 - 200 Hz were kept.

\begin{exmp}
    \textit{To illustrate how extracted features from the FSST impact on the choice of the input layer size, consider a framed signal by the method in \ref{subsec:framing} yielding a patch length of 2000. As Figure \ref{fig:workflow} suggests, we have to extract the features from the given patches, but for the sake of this example we take into account only one patch. The FSST will compute time-frequency characteristics of the given patch providing a fixed amount of instantaneous frequencies. For instance, we will get a feature matrix $F \in \mathbb{C}^{q \times p}$}, where $q$ is number of computed frequencies and $p$ the number of timestamps (the later matches the length of the patch). However, we might be interested in a range of frequencies, and by selecting those, we get a smaller fixed number of frequencies that we can feed the input layer with, and so the architecture of the input layer should be properly configured.
\end{exmp}

\subsection{Neural Network}\label{subsec:neural-network}

The proposed approach in this work is DRNNs specialized in time-series data.
Knowledge about past and future times is a great feature to have for neural networks and Long Short-Term Memory (LSTM) excel in this subject.
One shortcoming in RNNs is the \textit{vanishing gradient problem}. 
Networks that are deep and have a large number of units tend to vanish the gradient when \textit{Backpropagation Through Time} (BPTT) algorithm is computed.
So LSTM has proved to be a robust solution to the vanishing gradient problem and, at the same time, keeping the well-known benefits of RNNs.

\subsubsection{Long Short-Term Memory (LSTM) Block} \label{subsubsec:lstm-block}

\begin{figure}[H]
    \centering
    \includegraphics[scale=0.41]{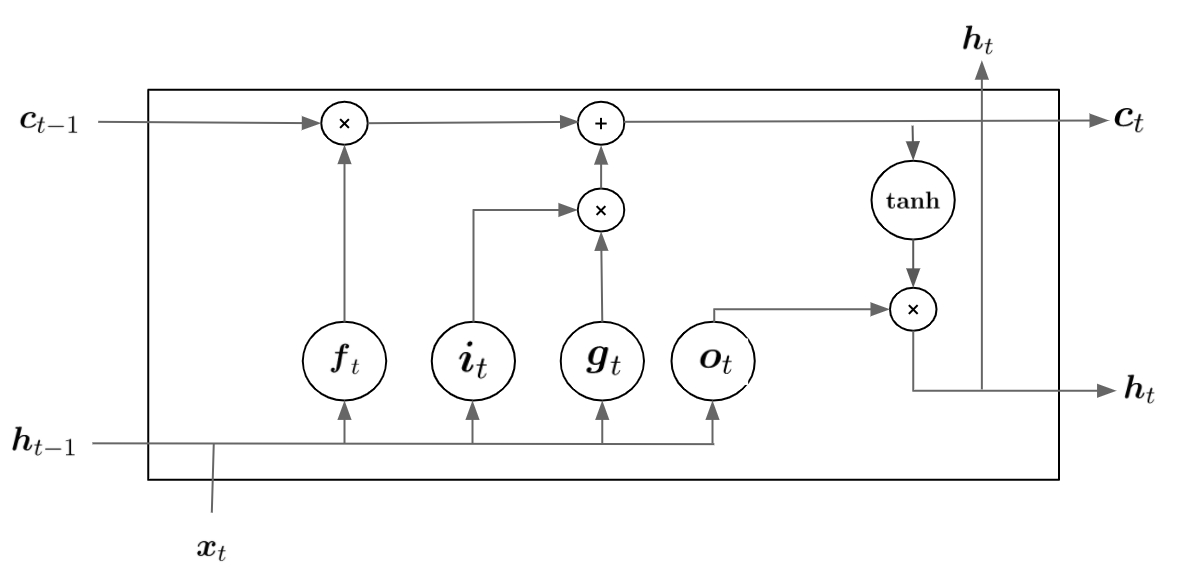}
    \caption{Fundamental block in LSTM networks}
    \label{fig:lstm-block}
\end{figure}

The LSTM network is a type of RNNs.
It consists of several memory blocks. 
Figure \ref{fig:lstm-block} reflects the operations taking place within each one of them at a specific layer.

For a given block and features $\boldsymbol{x}_t$ at a given time $t$ in an LSTM layer, $l$, the following equations are computed:

\begin{equation}
    \boldsymbol{i}^{(l)}_t = \sigma(\boldsymbol{W}^{(l)}_{ii} \boldsymbol{x}^{(l)}_t + \boldsymbol{b}^{(l)}_{ii} + \boldsymbol{W}^{(l)}_{hi}\boldsymbol{h}^{(l)}_{t-1} + \boldsymbol{b}^{(l)}_{hi})
\end{equation}

\begin{equation}
    \boldsymbol{f}^{(l)}_t = \sigma(\boldsymbol{W}^{(l)}_{if} \boldsymbol{x}^{(l)}_t + \boldsymbol{b}^{(l)}_{if} + \boldsymbol{W}^{(l)}_{hf}\boldsymbol{h}^{(l)}_{t-1} + \boldsymbol{b}^{(l)}_{hf})
\end{equation}

\begin{equation}
    \boldsymbol{g}^{(l)}_t = \mathbf{tanh}(\boldsymbol{W}^{(l)}_{ig} \boldsymbol{x}^{(l)}_t + \boldsymbol{b}^{(l)}_{ig} + \boldsymbol{W}^{(l)}_{hg}\boldsymbol{h}^{(l)}_{t-1} + \boldsymbol{b}^{(l)}_{hg})
\end{equation}

\begin{equation}
    \boldsymbol{o}^{(l)}_t = \sigma(\boldsymbol{W}^{(l)}_{io} \boldsymbol{x}^{(l)}_t + \boldsymbol{b}^{(l)}_{io} + \boldsymbol{W}^{(l)}_{ho}\boldsymbol{h}^{(l)}_{t-1} + \boldsymbol{b}^{(l)}_{ho})
\end{equation}

\begin{equation}
    \boldsymbol{c}^{(l)}_t = \boldsymbol{f}^{(l)}_t \odot \boldsymbol{c}^{(l)}_{t-1} + \boldsymbol{i}^{(l)}_t \odot \boldsymbol{g}^{(l)}_t
\end{equation}

\begin{equation} \label{eq:hidden-state}
    \boldsymbol{h}^{(l)}_t = \boldsymbol{o}^{(l)}_t \odot \mathbf{tanh}(\boldsymbol{c}^{(l)}_t)
\end{equation}

Where in Figure \ref{fig:lstm-block}, $\boldsymbol{h}_{t}$ is the hidden state at time $t$, $\boldsymbol{c}_{t}$ is the cell state at time $t$, $\boldsymbol{x}_{t}$ is the input at time $t$, $\boldsymbol{h}_{t-1}$ is the hidden state of the layer $t-1$ or the initial hidden state at time 0, and $\boldsymbol{i}_{t}$, $\boldsymbol{f}_{t}$, $\boldsymbol{g}_{t}$, $\boldsymbol{o}_{t}$ are the input, forget, cell and output gates, respectively. $\sigma$ is the sigmoid function, and $\odot$ is the Hadamard product.

In a multi-layer scheme the hidden state $\boldsymbol{h}^{(l-1)}_t$ of a previous layer can be multiplied by a drop-out $\delta^{(l-1)}_t$ coefficient where each $\delta^{(l-1)}_t$ is a random Bernoulli variable with probability $p$.

LSTM cells decide whether to keep information from a previous time $t-1$ using the forget gate by taking into account $\boldsymbol{x}_t$ and $\boldsymbol{h}_{t-1}$. 
Then using the input gate and the cell gate can choose what information is a candidate to be stored in the cell.
Once the information has been chosen $\boldsymbol{c}_{t-1}$ is updated into the new state $\boldsymbol{c}_t$.
Finally, the hidden state $\boldsymbol{h}_t$ is computed in \Cref{eq:hidden-state}.

\subsubsection{Bidirectional Long Short-Term Memory (BiLSTM)}

\begin{figure}[H]
    \centering
    \includegraphics[scale=0.475]{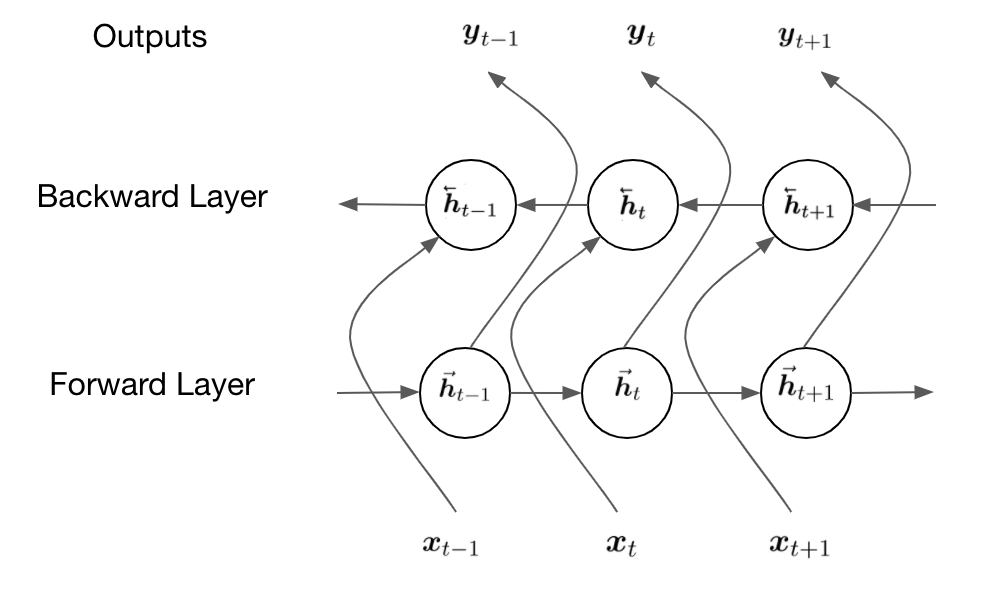}
    \caption{BiLSTM network}
    \label{fig:BiLSTM}
\end{figure}

BiLSTM networks are an extension to LSTM, in which training is performed in both time directions simultaneously possible by using two embedded RNN layers. 
One backward and another one forward depicted in Figure \ref{fig:BiLSTM}.
Both backward and forward hidden-states are then fed into the next layer. In some instances, it can be another BiLSTM layer.
It is worth mentioning that each block comprises the computations described in Section \ref{subsubsec:lstm-block}.

Bidirectional Recurrent Neural Networks (BRNN) can also be built upon different RNN schemes such as Bidirectional Gated Recurrent Units (BGRU).

\subsubsection{Architecture}

\begin{figure}[H]
    \includegraphics[scale=0.45]{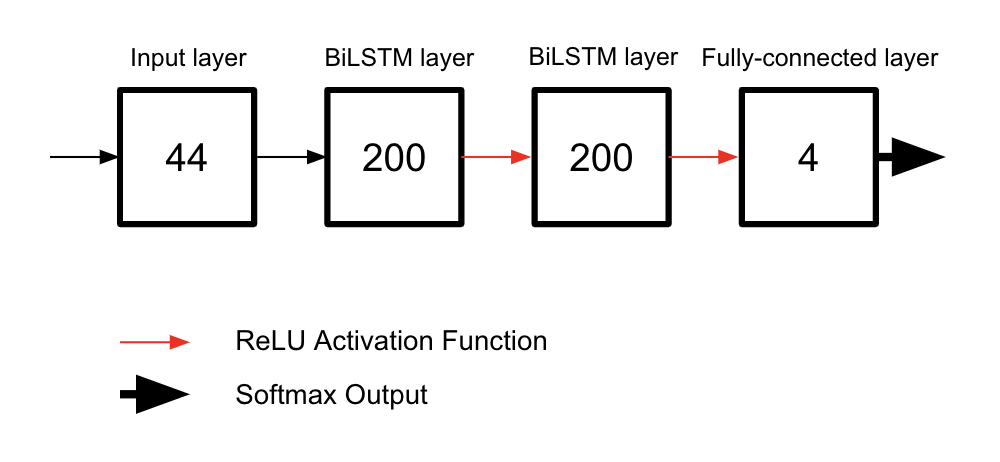}
    \caption{Proposed DRNN.}
    \label{fig:architecture}
\end{figure}

The LSTM architecture in Figure \ref{fig:architecture} is comprised of three hidden layers besides the input and output layer.
The input layer has a dimensionality of 44 correspondings to the features of interest.
It is connected to a first 200-unit BiLSTM hidden layer activated by a ReLU function which is also connected to a second hidden layer with the same characteristics.
Its output is then fed into a fully-connected or dense layer to compute the corresponding scores.
Finally, a softmax layer is liable for computing the likelihood of belonging to a particular class.
This architecture could seem straightforward and paltry, but it does the job classifying with a more than worthy performance.


\subsubsection{Drop-out Regularization}

Deep neural networks usually tend to be very deep. 
Convolutional Neural Networks (CNN) is one good example. 
Networks having a large number of hidden layers yields to have a large number of parameters to optimize.
Nevertheless, the more layers a DNN has, the more likely it is to overfit. 
Suggesting that it will learn every detail off of the data trained with and it will lose generality.

Drop-out is a technique for addressing this problem.
The central idea is to randomly drop units from the network during training and preventing neurons from co-adapting too much. 
In broad, drop-out is implemented by deactivating neurons in each iteration with a probability $p$. 
During the whole training process, all neurons will have been deactivated the same amount of times.

This technique significantly reduces overfitting and gives significant improvements over other regularization methods.

\subsection{Model training \& testing}\label{subsec:training}

Training was performed using cross-validation techniques, namely \textit{K-Fold Cross Validation}.
Training.
Validation and test folds were deliberately chosen to begin training the neural network. 
With K being the number of folds which was picked to be 10.
Along these lines, every observation was used to train and test the model. 
Say that the picked model is the one performing at its best on the testing dataset.

Most of the time contributed in training is used to choose the appropriate hyperparameters of the model.
Generally, this is done by trial and error.
The chosen hyperparameters that best performed are shown in Table \ref{tab:hyperparameters}.

A gradient threshold is defined to avoid any issues with gradient explosions.
Therefore, if the gradient in absolute value is greater than the threshold, it will be clipped.
Another technique to avoid overfitting is using a validation set to control how close the validation loss is to the training loss called \textit{Early Stopping}.
The stopping is set by defining a \textit{validation patience} which outlines how many times the validation loss can be higher than the minimum validation loss computed at a given iteration.
If this criterion is met, then the training progress stops.

The optimizer is another option to set up.
\textit{Adaptive Moment Estimation} (ADAM) is a stochastic gradient-based optimization method to find the weights in a neural network. It requires that a learning rate is set and it can also be adaptive.
It implies that every a fixed amount of epochs the learning rate decreases by some factor.

The hyperparameters in Table \ref{tab:hyperparameters} were chosen with certain criterion. For instance, the mini-batch size, initial learning rate, the learn rate drop period and gradient threshold are the most common values in the literature, which were found to have the best results based on the architecture and data selected. Ultimately, the number of epochs generally are defined between 10-30, although via trial and error, after 6 epochs we noticed the performance of did not improve whatsoever if the network was trained for longer epochs, and in some cases, the network was prone to overfit. Thereby, 6 epochs seems a reasonable value to reduce the training time and the possibility of overfitting. Addtionally, a validation patience was added, and in this case chosen to be 6 because we have seen that the accuracy did not improve after 6 failures.

\begin{table}[H]
    \centering
    \renewcommand{\arraystretch}{1.3}
    \caption{\textsc{Chosen Hyperparameters}}
    \begin{tabular}{c c} 
    \toprule[1.5pt]
    Hyperparameter & Value \\
    \midrule[1.5pt]
    Optimizer & ADAM \\
    Epochs & 6 \\
    Mini-Batch Size & 50 \\
    Initial Learning Rate & 0.01 \\
    Learn Rate Drop Period & 3 \\
    Gradient Threshold & 1 \\
    Validation Patience & 6 \\
    \bottomrule[1.5pt]
    \end{tabular}
    \label{tab:hyperparameters}
\end{table}


\section{Results}\label{sec:results}

Performance of a model is a crucial step to select the appropriate model. 
Most common metrics reported are precision ($P_+$), sensitivity (Se), F1-score ($F_1$) and accuracy (ACC) computed based on true positives (TP), true negatives (TN), false positives (FP) and false negatives (FN).

\begin{equation} \label{eq:acc}
    ACC = \frac{TP+TN}{TP+TN+FP+FN}
\end{equation}

\begin{equation} \label{eq:precision}
    P_+ = \frac{TP}{TP+FP}
\end{equation}

\begin{equation} \label{eq:sensitivity}
    Se = \frac{TP}{TP+FN}
\end{equation}

\begin{equation} \label{eq:f1}
    F_1 = 2\frac{P_+ \cdot Se}{P_+ + Se}
\end{equation}

\begin{figure}[H]
    \centering
    \includegraphics[scale=0.215]{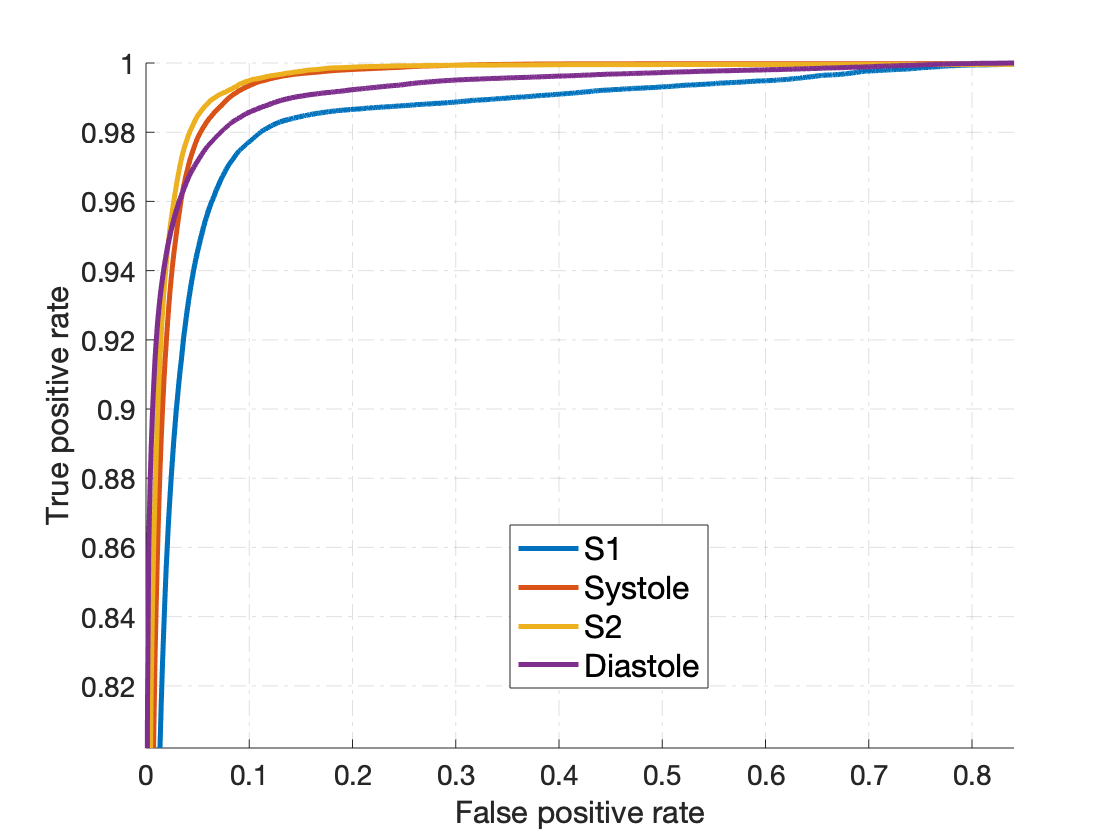}
    \caption{Receiver Operating Characteristic curves. Curves corresponding to each class (S1, sys, S2, dias).}
    \label{fig:roc-curve}
\end{figure}

Another necessary result to report is the so-called ROC curve, computed off the scores from the last hidden layer and test labels.
In Figure \ref{fig:roc-curve}, a ROC curve is depicted for a given fold, which is constructed by plotting True Positive Rate (TPR) and False Positive Rate (FPR). TPR is also known as sensitivity defined in \Cref{eq:sensitivity} and FPR is also known as \textit{fall-out} formulated in \Cref{eq:fall-out}.

\begin{equation} \label{eq:fall-out}
    FPR = \frac{FP}{FP + TN}
\end{equation}

Once the ten models have been trained, one has to be selected. 
\textit{Area Under the Curve} (AUC) is computed for each model in Table \ref{tab:auc-scores}, and the highest is picked. 
Since there are four classes, the average is computed and used for comparison.

\begin{table}[H]
    \centering
    \renewcommand{\arraystretch}{1.3}
    \caption{\textsc{AUC Scores}}
    \begin{tabular}{c c} 
    \midrule[1.5pt]
    K-Fold & AUC (\%) \\
    \toprule[1.5pt]
    1 & 99.1 \\
    2 & 98.5 \\
    3 & 98.5 \\
    4 & 97.2 \\
    5 & 98.9 \\
    6 & 98.1 \\
    7 & 98.5 \\
    8 & 98.9 \\
    9 & 98.9 \\
    10 & 98.9 \\
    \bottomrule[1.5pt]
    \end{tabular}
    \label{tab:auc-scores}
\end{table}

Lastly, \Crefrange{eq:acc}{eq:f1} are averaged across all trained models to report the final results in Table \ref{tab:results}.

\begin{table}[H]
    \centering
    \renewcommand{\arraystretch}{1.3}
    \caption{\textsc{Final results upon cross-validation.}}
    \begin{tabular}{c c c c c} 
    \toprule[1.5pt]
    State & $P_+$ (\%) & Se (\%) & $F_1$ (\%) \\
    \midrule[1.5pt]
    S1 & 85.7 & 86.5 & 86.0 \\
    Sys & 90.0 & 90.0 & 90.0 \\
    S2 & 87.1 & 86.7 & 86.9 \\
    Dias & 94.9 & 94.6 & 94.8 \\
    \midrule[1pt]
    Average & 89.3 & 89.5 & 89.4 \\
    \midrule[1pt]
    ACC (\%) & 91.34 & & \\
    \bottomrule[1.5pt]
    \end{tabular}
    \label{tab:results}
\end{table}

\begin{table}[H]
    \centering
    \renewcommand{\arraystretch}{1.3}
    \caption{\textsc{State-of-the-art algorithms comparison.}}
    \begin{tabular}{c c c c c c} 
    \toprule[1.5pt]
    Algorithm & ACC (\%) & $P_+$ (\%) & Se (\%) & $F_1$ (\%) \\
    \midrule[1.5pt]
    BiLSTM & 91.34 & 89.3 & 89.5 & 89.4 \\
    LR-HSSM \cite{7234876} & 92.52 & 95.92 & 95.34 & 95.63 \\
    CNN+HMM \cite{8620278} & 93.7 & 95.7 & 95.7 & 95.7 \\
    \bottomrule[1.5pt]
    \end{tabular}
    \label{tab:algorithms-comparison}
\end{table}

\section{Conclusion}\label{sec:conclusion}

In this work, a novel heart sounds segmentation model has been presented.
An LSTM model is accompanied by a time-frequency feature extraction procedure carried out by the Fourier Synchrosqueezed Transform (FSST).
It is shown that choosing the right method to extract features and the right neural network architecture yields, in comparison to other proposals, an almost state-of-the-art performance as shown in Table \ref{tab:algorithms-comparison}.
For instance, the modified U-Net neural network used by Renna \textit{et al.} \cite{8620278} with more than 20 layers.

Data augmentation and data addition can also be increased to make the network deeper and to train it for a more significant number of epochs. 
This means that using an LSTM network can achieve higher performances by developing a more complex architecture since these cells are specialized in time-series data.

A possible future extension from this approach is to implement a post-processing stage in order to imply correct transitions between states, which is the main limitation of this approach. Thus, improving the performance.



\bibliographystyle{ieeetran}
\bibliography{elektron}

\end{document}